\documentclass{aa}
\usepackage{graphicx}
\usepackage{txfonts}
\usepackage{xcolor}
\usepackage[normalem]{ulem}

\begin{document} 

   \title{Disentangling Milky Way halo populations at low metallicities using [Al/Fe]\thanks{Based on observations collected at the European Southern Observatory under ESO programme IDs 109.22 and 110.240 (PI: Á. Skúladóttir) as well as obtained from the ESO Science Archive Facility under the ESO programme 8.B-0475, 076.B-0133, 090.B-0605, 109.23, 71.B-0529, and 68.D-0094}}
   \subtitle{}

   \author{H. Ernandes\thanks{\email{heitor.ernandes@geol.lu.se}}
          \inst{1}
          \and
          Á. Skúladóttir\inst{2}
          \and
          S. Feltzing\inst{1}
          \and
          D. Feuillet\inst{3}
          }

   \institute{Lund Observatory, Department of Geology, S\"olvegatan 12, SE-223 62 Lund, Sweden
   \and
   Dipartimento di Fisica e Astronomia, Universitá degli Studi di Firenze, Via G. Sansone 1, I-50019 Sesto Fiorentino, Italy
   \and 
   Observational Astrophysics, Department of Physics and Astronomy, Uppsala University, Box 516, SE-751 20 Uppsala, Sweden
              }

   \date{Received 8 September 2025 / Accepted 26 September 2025}

 
  \abstract
{Differentiating between in situ and accreted populations in the Milky Way halo is a challenging task. Various kinematic spaces are often used to identify distinct accreted populations from the in situ Milky Way halo. However, this approach has limitations, especially at low orbital energies. To overcome this ambiguity, elemental abundances are typically used to distinguish between populations. However, for many elemental abundance ratios (e.g. [Mg/Fe]), it remains difficult to make this distinction at low metallicities ([Fe/H]$\sim-$1.5). Aluminium abundances, on the other hand, are empirically effective discriminators and separate accreted from in situ populations in the Milky Way halo, even at low metallicities and low orbital energies.}
   {We aim to test the discriminating power of [Al/Fe] using a well-studied sample of high-velocity stars in the solar vicinity with high-quality spectra. With these stars, we explore the ability of [Al/Fe] to separate in situ from accreted stars and test its limitations.}
   {We derived aluminium abundances from the Al I 3944 and 3961 {\rm \AA} lines for 45 stars observed in two ESO programmes, along with 11 stars with archival spectra. Aluminium abundances were determined using 1D LTE and 1D NLTE spectral synthesis and line-profile fitting.}
   {We confirm that the halo low-$\alpha$ population (associated with accreted stars) systematically exhibits lower [Al/Fe] compared to halo high-$\alpha$ (in situ) stars. However, at low $\rm[Fe/H]\approx-1.4$, we identify three stars, previously classified as in situ or thick-disc stars, as having low $\rm[Al/Fe]<-0.3$, indicating that they might actually be accreted.
  }
   {Aluminium abundances, when carefully measured and with NLTE effects taken into account, are effective tracers of the chemical history of halo stars within the [Fe/H] limits analysed in the present work ($-1.5\lesssim$[Fe/H]$\lesssim -0.5$). They provide an independent constraint on origin, complementing $\alpha$-element abundance trends, and help disentangle subpopulations within the accreted halo, especially in the metal-poor regime.}


\maketitle
%

\section{Introduction}

The stellar halo of the Milky Way preserves the signatures of its hierarchical assembly. Early studies revealed that the Galactic halo is not chemically homogeneous but instead contains stars with distinct enrichment histories. By studying halo stars passing through the solar neighbourhood, \citet{Nissen10} discovered that they fall into two chemically distinct groups: a high-$\alpha$ and a low-$\alpha$ population. With the advent of {\it Gaia} \citep{Gaia16b}, it has become clear that the Galactic halo shows evidence of multiple accretion events, the largest of which is the {\it Gaia}–Sausage–Enceladus (\citealt{Helmi18, Belokurov18}). The low-$\alpha$ population identified by Nissen \& Schuster has since been shown to largely correspond to this accreted dwarf galaxy \citep{Haywood18,Helmi18}. Recently, analysis of low-$\alpha$ stars has provided new insights into how this accretion event occurred \citep[][]{Asa25}.  

There are several ways to disentangle accreted stars from the in situ halo population. One option is kinematic selection, which uses orbital actions and their angular momenta to isolate debris from the in situ components \citep[e.g.][]{Feuillet20}. However, purely kinematic criteria vary in completeness and purity, depending on the adopted potential and selection method \citep{Carrillo23, Feltzing23,Ernandes24}. Another approach combines chemical selection with kinematics, using elemental abundances to distinguish populations \citep[e.g.][]{Limberg22, Buder21, Naidu22}. In this context, aluminium abundances have recently emerged as a particularly powerful tool. Empirical studies have shown that accreted stars and stars in dwarf galaxies display systematically lower [Al/Fe] than in situ stars at the same metallicity \citep{Hawkins15,Hasselquist21,Feuillet22}. Moreover, aluminium abundances have been shown to separate not only in situ from accreted halo stars, but also distinct subpopulations within the accreted component \citep{Asa25}.  

The nucleosynthetic origin of aluminium helps to explain its discriminating power. Aluminium is primarily produced in massive stars during hydrostatic carbon burning, with additional contributions from neon burning \citep{WW95}. Under these conditions, it behaves as a primary element and shows little dependence on metallicity. However, secondary production channels, involving $\alpha$-, n-, and p-capture reactions on neutron-rich seed nuclei, such as $^{22}$Ne and $^{23}$Na, also contribute \citep{RoedererLawler21}. As a result, aluminium can exhibit metallicity-dependent behaviour. Indeed, \citet{Kobayashi20} point out that in the Milky Way, aluminium follows trends characteristic of a secondary element. Importantly, the relative efficiency of these processes depends on the initial mass function, star-formation history, and chemical-enrichment timescales, which differ between the Milky Way and its satellites. This dependence of aluminium production efficiency on various parameters of galaxy evolution explains why aluminium abundances can vary systematically between accreted and in situ populations.  

The Nissen \& Schuster sample provides an ideal dataset to test the ability of aluminium to distinguish halo populations. The stars were selected to have high space velocities ($V_{\rm total} > 180$\,km\,s$^{-1}$) relative to the local standard of rest, ensuring a high probability of belonging to the halo. Additional selection criteria used the Strömgren indices to select dwarfs and subgiants with 5200\,K $< T_{\rm eff} <$ 6300\,K and [Fe/H] $> -1.6$, to reduce parameter-dependent systematic effects. Because of this narrow parameter range, non local thermodynamic equilibrium (NLTE) corrections are expected to be relatively uniform across the sample. The original discovery of the high- and low-$\alpha$ sequences was based on this dataset, which has since become a benchmark for studies of the Galactic halo.  

An important feature of the Nissen \& Schuster sample is that it provides small-number statistics with very high precision \citep{LindegrenFeltzing13}. This means that, although modest in size, the dataset offers some of the most precise stellar parameters and elemental abundances available. Until recently, aluminium abundance could not be derived for these metal-poor stars because of the lack of observations of the near-UV lines. With new observations, together with archival UVES spectra covering the bluer wavelength region, it is now possible to measure aluminium using its resonance lines at 3944 and 3961 {\rm \AA} in this key sample for the first time.  

The aim of this work is to test the capability of aluminium abundances to separate in situ from accreted halo stars, using the Nissen \& Schuster sample as a benchmark. We explore the potential of aluminium as a substructure tracer within the accreted halo in \cite{Asa25}, and here we discuss both its discriminating power and its limitations.

\section{Observations, archival data, and data reductions}

\subsection{Target selection and data reduction}

The stellar sample analysed in this work is a subsample of the 66 stars studied by \citet{Nissen10, Nissen11, Nissen24}. From this larger set, we selected stars that could be observed in the Southern hemisphere with VLT/UVES \citep{Dekker00}.

New spectra were obtained for 45 stars with the blue arm of UVES. These data were collected in two ESO programmes (PI: Á. Skúladóttir, ESO ID 109.22 and 110.240). The observations were carried out at high spectral resolution ($R > 40{,}000$) and achieved a signal-to-noise ratio greater than 80 per pixel at 4000{\rm \AA}. We used the UVES 390 setting, which covers the 3260–4540\,{\rm \AA} wavelength range \footnote{https://www.eso.org/sci/facilities/paranal/instruments/uves/doc.html}.  

In addition to the new observations, we included 11 stars with suitable archival UVES spectra. These data also cover the region of the Al\,I lines (3900–4000\,{\rm \AA}) with similarly high signal-to-noise ratios (${\rm S/N} > 80$\,pix$^{-1}$ at 4000\,{\rm \AA}). The archival material comes from ESO programmes 68.B-0475, 076.B-0133, 090.B-0605, 109.23, 71.B-0529, and 68.D-0094.  

Both the new and archival spectra were reduced using the standard UVES pipeline \citep{Freudling13}. Most stars were observed with the UVES 390 setting, but three stars — G188-22, HD~25704, and G53-41—  were observed with the 437 setting, which covers 3730–4990\,{\rm \AA} and therefore includes the Al\,I lines.


\begin{figure*}
    \centering
    \includegraphics[width=0.95\linewidth]{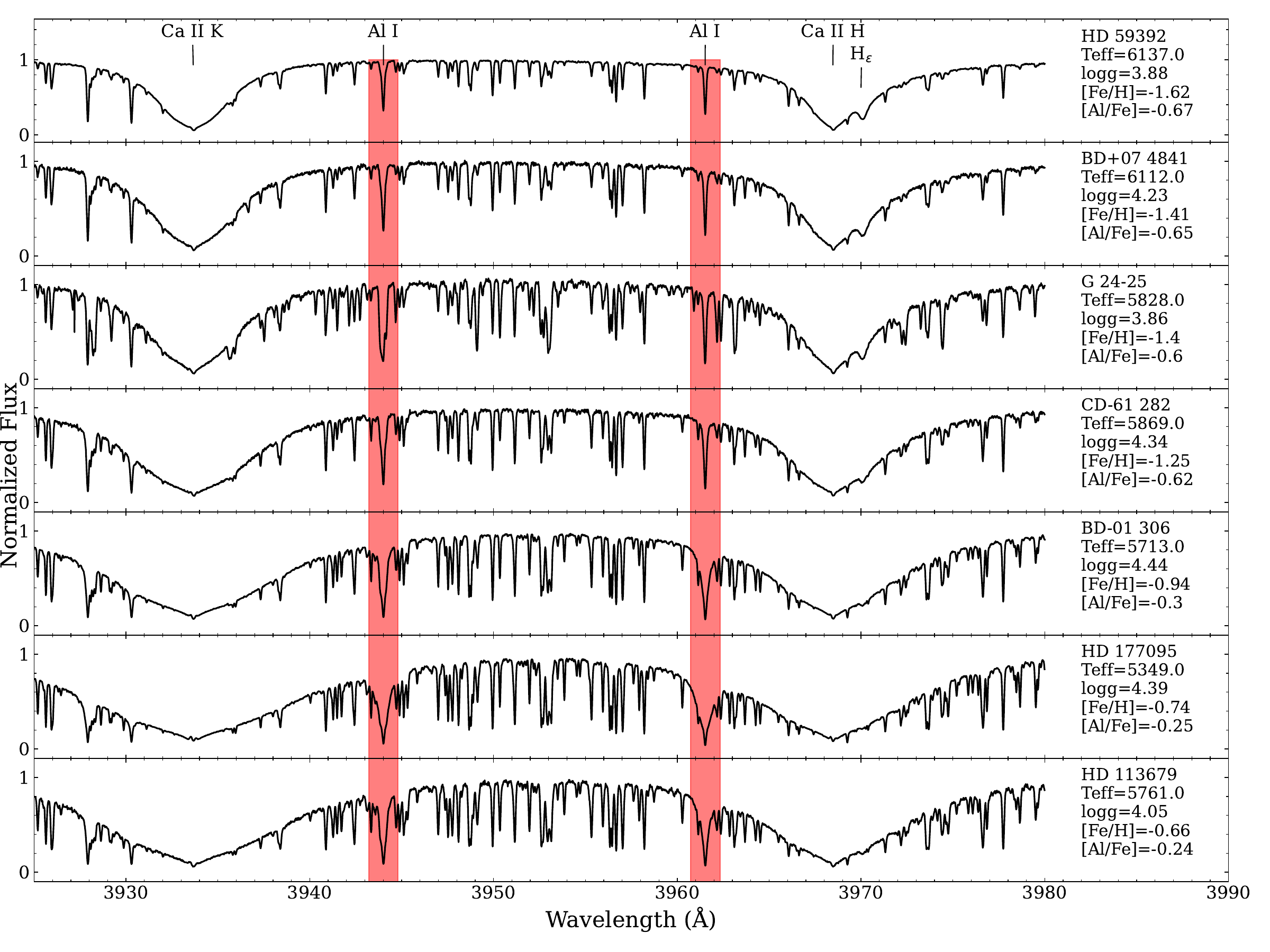}
    \caption{Normalised spectra for seven stars spanning a range of metallicities, from $\rm [Fe/H]=-1.41$ at the top to $\rm [Fe/H]=-0.66$ in the bottom panel. The Al I 3944 and 3961 {\rm \AA} lines used to derive Al abundance are highlighted in red. The position of the Ca\,II H \& K lines and H$_\epsilon$ are also indicated.}
    \label{spectra}
\end{figure*}

\subsection{Radial velocity}

Following data reduction and application of the wavelength solution, all spectra were corrected for heliocentric velocity. This step ensures that the observed wavelengths are placed in the rest frame of the Sun, which facilitates consistent comparison between the observations.

Radial velocities were determined using a minimisation procedure in which synthetic spectra, computed with \textsc{Turbospectrum} \citep{Turbo98,Turbo12}, were fitted to the observations. The stellar parameters adopted for the synthetic spectra were taken from \citet{Nissen24}. By comparing each observed spectrum with its corresponding synthetic template, we identified the velocity shift that minimised the residuals, thereby providing a robust measurement of the stellar radial velocity.  

For each star, the median radial velocity derived from all available observations is reported in Table~\ref{tab:rv}. This approach reduces the impact of noise or possible systematics in individual exposures. We note that the star HD~111980 is a single-lined spectroscopic binary (SB1) and that its radial velocity variability is therefore intrinsic to its binary nature.

\subsection{Co-addition and normalisation}

After applying the radial velocity correction to each exposure, we co-added the spectra for each star by summing the unnormalised flux across the UVES blue and red arms (settings 390, 437, 554l, and 564u). This procedure preserves the flux information and ensures that the signal-to-noise ratio in each region is maximised. Some stars were observed up to nine times (see Table~\ref{tab:rv}), whereas most have only one exposure. In this work, we focus on the 390 and 437 settings, which were used to derive the aluminium abundances.  

Normalisation of the spectra was performed using a method based on continuum points determined from a synthetic spectrum template created with \textsc{Turbospectrum}. Because this approach incorporates information about line positions and broadening, it provides a reliable identification of the true continuum. A three-step procedure was implemented to mitigate observational artefacts, such as telluric absorption, cosmic rays, and residual signatures of the blaze function in individual echelle orders. In the first step, a renormalisation placed the synthetic and observed spectra on the same overall scale while preserving their shapes. The second step identified continuum points in the observed spectra based on the locations predicted by the synthetic template. Finally, a third pass re-evaluated the continuum points in the processed observations and refined the continuum placement.  

For the 390 setting, this procedure was applied separately in three spectral regions to account for the challenges posed by the near-UV, particularly around the Balmer jump. A template atmosphere with parameters $T_{\mathrm{eff}} = 5900$\,K, $\log g = 4.2$, [Fe/H] = $-1.0$, and $v_{\mathrm{mic}} = 1.0$\,km\,s$^{-1}$ was used to identify candidate continuum points. Points above a threshold flux of 0.99 were retained and their analogues in the observed spectra were fitted with a third-order Chebyshev polynomial. After the first normalisation, this procedure was repeated on the observed spectra, selecting only continuum points that were both above the threshold and present in the initial set. A final polynomial fit was then adopted as the continuum. The same approach was applied to the 437, 564u, and 564l settings, without requiring splitting into multiple regions. The 564u and 564l settings will be analysed in Shejeelammal et al. in prep.

Examples of the resulting spectra are shown in Fig.~\ref{spectra}. These illustrate the particular difficulties of continuum placement in the near-UV, where the Ca\,II H and K lines dominate the spectral region. The challenge becomes especially pronounced at higher metallicities, where line crowding is severe (bottom spectra in Fig.~\ref{spectra}).

\section{Atomic data}
The atomic data used in the elemental abundance analysis are provided in Table \ref{tab:atom}, including references for the energy levels and the log($gf$). We used VALD as the source of the hyperfine structures (hfs)\footnote{https://vald.astro.uu.se/}, with the original sources of the hyperfine structures given by \citet{hfs1BPM,hfs2SLa,hfs3BE,hfs4JLS}.

\begin{table*}
    \centering
\caption{Atomic data for the Al I and Ca II lines analysed.}
\resizebox{0.95\textwidth}{!}{ \begin{tabular}{lccccccccccc}
\hline
\hline
 Species & Wavelength &  $\chi_{ex}$ & log$(gf)$ &  \multicolumn{3}{c}{Lower Level} & \multicolumn{3}{c}{Upper Level}  & Reference \\
     &      &              &         & Energy & ConFig.& J & Energy & ConFig.& J & \\
     &  [\rm \AA]    &     [eV]         &         & [cm$^{-1}$] & &  & [cm$^{-1}$] &  &  & \\    
\hline
   Al I   &  3944.01  & 0.000 & -0.623 & 0.000 & 3s$^{2}$3p $^{2}$P$^{o}$ & 1/2 & 25347.756  & 3s$^{2}$4s $^{2}$S & 1/2 & \cite{Eriksson61AL} \\
   Al I   &  3961.52  & 0.014 & -0.333 & 112.061 & 3s$^{2}$3p $^{2}$P$^{o}$ & 3/2 & 25347.756 & 3s$^{2}$4s $^{2}$S & 1/2 & \cite{Eriksson61AL}  \\
   \hline
   Ca II   & 3933.66  &  5.884  & 0.135 & 0.000 & 3p$^{6}$4s $^{2}$S & 1/2 & 25414.40 & 3s$^{6}$4p $^{2}$P$^{o}$ & 3/2  &  \cite{Edlen56Ca}\\
   Ca II  & 3968.47   &  4.131  & -0.18 & 0.000 & 3p$^{6}$4s $^{2}$S & 1/2 & 25191.51 & 3p$^{6}$4p $^{2}$S & 1/2  & \cite{Edlen56Ca} \\
   
   \hline
   \hline
    \end{tabular}}
    \label{tab:atom}
\end{table*}

\section{Determination of aluminium abundances}

\subsection{Stellar atomospheric parameters}

The stellar parameters for most of our sample were adopted from \citet{Nissen24}, which were derived by \citet{Amarsi19pars}. For stars not included in that work, we used the parameters reported by \citet{Nissen11}, as indicated by the NSflag in Table~\ref{tab:pars}. The stellar parameters used here follow \citep{Nissen10}. Later work \citep[e.g.][]{Nissen24} updated these values using Gaia parallaxes and 3D-NLTE corrections, leading to small systematic shifts in effective temperatures, typically between 10 and 80 K (largest near the coolest turn-off temperatures), and yielding more precise surface gravities ($\sim \pm$0.05 dex).

For a subset of stars, 3D [Fe/H] values are available from \citet{Amarsi19,Amarsi22,Nissen24}. The difference between the 1D and 3D [Fe/H] values is small, with an average offset of only 0.02\,dex across our sample.  

For consistency in the abundance analysis, we adopted the 1D [Fe/H] values when computing the synthetic spectra. This choice ensures a uniform treatment across the full dataset.

\subsection{Aluminium abundances derived from the Al I 3944 and 3961 {\rm \AA} lines}

\begin{figure*}
    \centering
    \includegraphics[width=0.95\linewidth]{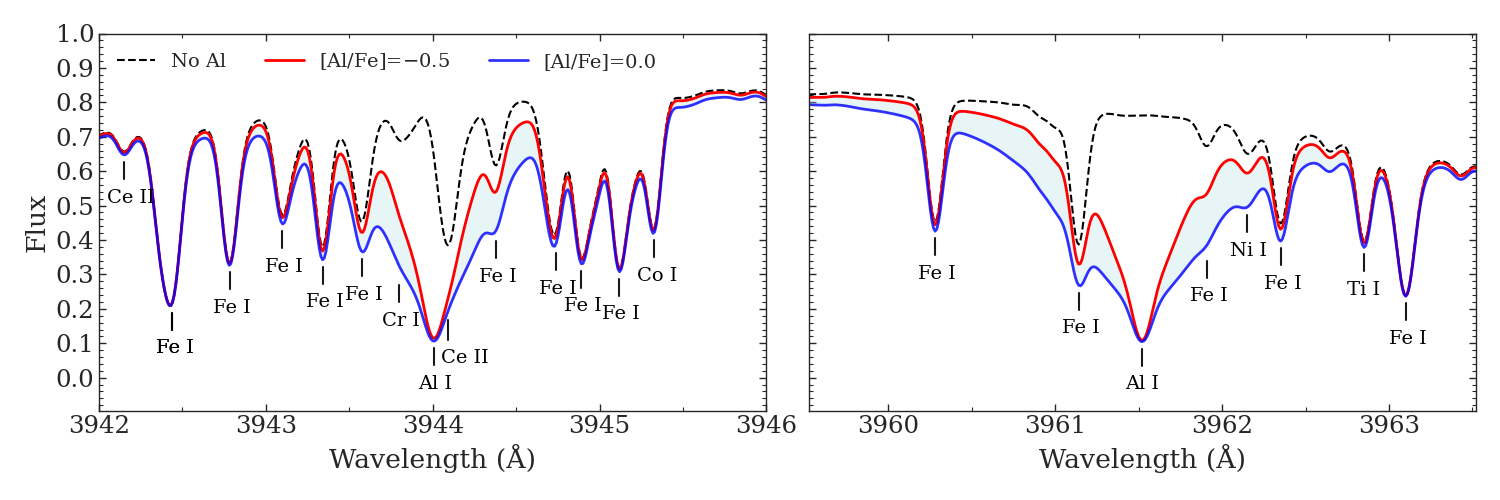}
    \caption{Synthetic NLTE spectra showing the shape of the aluminium lines 3944 {\rm \AA} and 3961 {\rm \AA} and their blends as indicated. For the synthetic spectra, the star HD 177095 at $\rm[Fe/H]=-0.74$ serves as an example for the NLTE calculations. Dashed lines indicate a spectrum with no aluminium present, while the red and blue lines show spectra with [Al/Fe]=$-0.5$ and $0.0$, respectively.}
    \label{fig:syn}
\end{figure*}

The aluminium abundances were derived from the resonance lines at 3944 and 3961\,{\rm \AA}. Both lines lie in the wings of the strong Ca\,II H and K features, within a spectral region that is heavily crowded with absorption lines, including the Balmer-$\epsilon$ (H$_\epsilon$) line (see Fig. \ref{spectra}). Because of this crowding, we fitted the aluminium lines individually using a non-automated approach, carefully reproducing their unique blend patterns, as illustrated in Fig.\ref{fig:syn}.

The 3961\,{\rm \AA} line is the least blended of the two (see Fig. \ref{fig:syn}) and was therefore adopted as the primary abundance indicator. Its main contamination arises from nearby Fe\,I lines. In contrast, the 3944\,{\rm \AA} line is affected by several blends: a Ce\,II line on the red wing, a  Cr\,II line on the blue wing, and additional Fe\,II lines. For this reason, the abundances derived from the 3944\,{\rm \AA} line are less reliable, although they were used to cross-check the results.  

Previous studies have shown that NLTE effects significantly impact the formation of the Al\,I lines. In particular, \cite{NordlanderLind17} demonstrate that for dwarf stars with metallicities in the range $-1.0 < \mathrm{[Fe/H]} < -0.5$, the NLTE correction for the 3961\,{\rm \AA} line is typically around +0.2\,dex in [Al/Fe] \citep[see also][]{Koutsouridou25}. These findings are supported by the empirical analysis of \citet{Roederer21}. Following these results, we performed abundance analyses in both NLTE and LTE. 

Synthetic spectra were computed with \textsc{Turbospectrum} \citep{Turbo98,Turbo12,TurboNLTE}, which provides a straightforward switch between the LTE and NLTE calculations. For the NLTE analysis, we employed the grids provided by \citet{Ezzeddine18}. Our fitting procedure first determined the broadening and calcium abundance that best reproduced the Ca\,II H and K wings, as these strongly influence the local continuum. The aluminium abundance was then derived by fitting the 3961\,{\rm \AA} line through $\chi^2$ minimisation, ensuring that both the line core and the extended wings were reproduced simultaneously. Finally, the abundance inferred from the 3944\,{\rm \AA} line was checked for consistency.  
Examples of the line profile, together with indications of the most prominent blending features, are shown in Fig.~\ref{fig:syn}.

The NLTE analysis also provides an improved reproduction of aluminium line profiles. In particular, the characteristic wing shape of the 3961\,{\rm \AA} feature is better matched under NLTE, supporting the reliability of the derived abundances. The typical NLTE correction for the 3961\,{\rm \AA} line is approximately +0.2\,dex, consistent with the predictions of \citet{NordlanderLind17} and the empirical findings of \citet{Roederer21}. Figure~\ref{stellar-plotnlte} illustrates these corrections as a function of stellar parameters, confirming that the magnitude of the effect depends mildly on the effective temperature but is largely insensitive to surface gravity.

\section{Uncertainties in the resulting abundances}

The uncertainties in the derived aluminium abundances arise primarily from the adopted stellar parameters. To estimate these effects, we followed the methodology of \citet{Ernandes20},\citet{Bensby04}, and \citet{Epstein_2010}, in which the atmospheric parameters are varied individually and the corresponding changes in the best-fit aluminium abundance are measured.  

For this sample, we adopted the typical parameter uncertainties reported by \citet{Nissen10}: $\Delta T_{\mathrm{eff}} = 50$\,K, $\Delta \log g = 0.10$\,dex, $\Delta v_{\mathrm{mic}} = 0.10$\,km\,s$^{-1}$, and $\Delta \mathrm{[Fe/H]} = 0.05$\,dex. Between \citet{Nissen10} and \citet{Nissen24}, the major improvement is the precision in $\log g$; however, since four of our stars rely on \citet{Nissen11} parameters, we adopted the more conservative, larger uncertainties to estimate our errors. We applied these variations to the 3961\,{\rm \AA} line and found abundance sensitivities of $(+0.05,\,-0.03,\,0.00, \text{and}\,-0.07)$\,dex for changes in $T_{\mathrm{eff}}$, $\log g$, $v_{\mathrm{mic}}$, and [Fe/H], respectively.  

We obtained the total uncertainty by summing these contributions in quadrature, yielding a typical error of 0.09\,dex in [Al/Fe] for the 3961\,{\rm \AA} line. This value represents the dominant source of random uncertainty in our measurements.  

Additional uncertainties may arise from continuum placement and line blending, especially for the 3944\,{\rm \AA} line. However, since we adopted the 3961\,{\rm \AA} line as the primary abundance indicator, these effects have a minimal impact on the final results, given the high quality of the data.

\section{Results: [Al/Fe] versus [Fe/H] trends}

\begin{figure}
    \centering
    \includegraphics[width=0.95\linewidth]{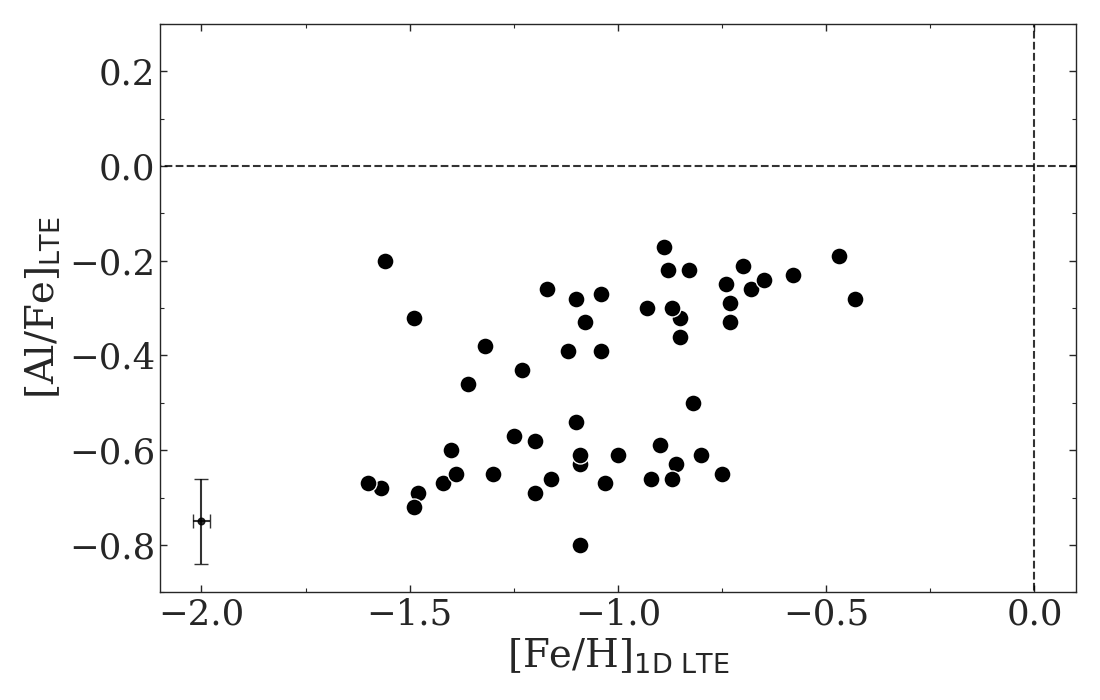}
    \includegraphics[width=0.95\linewidth]{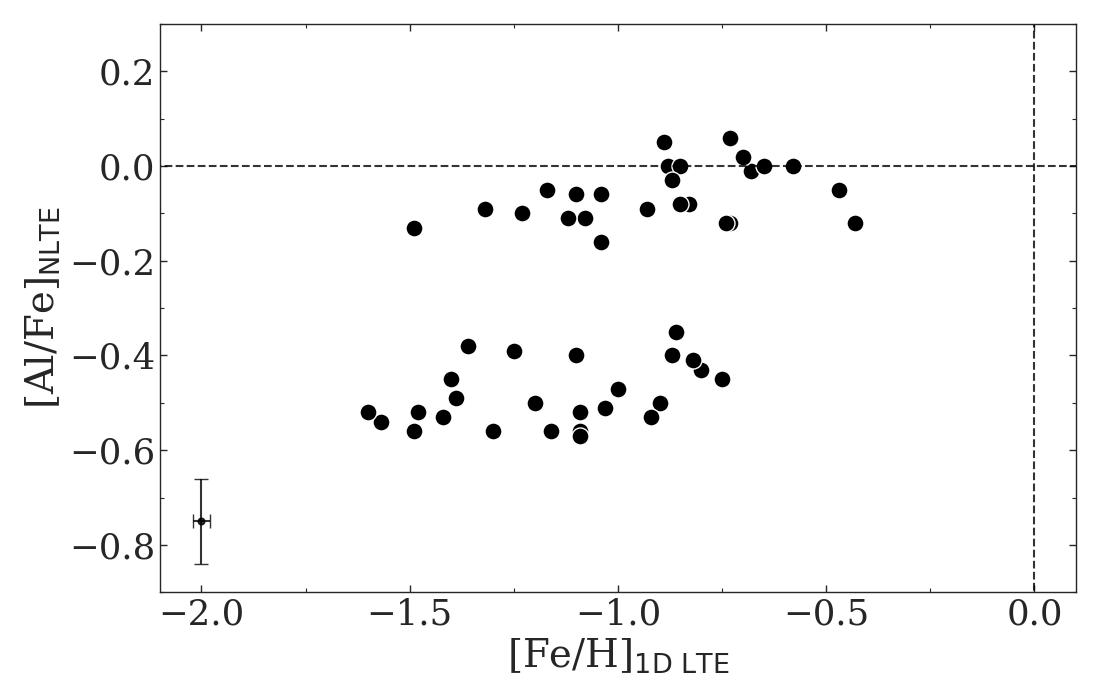}
    \caption{ Relation between [Al/Fe] and [Fe/H] for the full stellar sample, shown in LTE in the top panel and in NLTE in the bottom panel. The abundances are derived from the Al I 3961 {\rm \AA} line.}
    \label{Alplot}
\end{figure}

We measured aluminium abundances from both the 3944 and 3961\,{\rm \AA} lines and list the results in Table~\ref{tab:Abonds}. Figure~\ref{Alplot} shows [Al/Fe] as a function of [Fe/H] for the full stellar sample. The top panel presents the LTE abundances, while the bottom panel displays the abundances corrected for NLTE effects. In both cases, the results separate into two groups: low-Al and high-Al abundances. 

The separation between the two groups is already visible in LTE but becomes sharper in NLTE, demonstrating the importance of applying NLTE corrections when studying aluminium in metal-poor halo stars. A 3D NLTE analysis can change the shape of the Al\,I line near the core, as shown in Fig. 3 of \citet{NordlanderLind17}. However, \textbackslash{}citet\{NordlanderLind17\} note that both 1D and 3D models reproduce the line shape well after deriving the calcium abundance to set the continuum.

\begin{table}[]
    \centering
\caption{Aluminium and calcium abundances for the sample stars.}
\resizebox{0.45\textwidth}{!}{    \begin{tabular}{lcccccccccc}
\hline
\hline
 Star-id  & \multicolumn{2}{c}{[Al/Fe]$_{LTE}$} & $\Delta$[Al/Fe]$_{NLTE}$ & [Ca/Fe]$_{LTE}$ \\
 & 3944 \rm \AA & 3961 \rm \AA  & 3961 \rm \AA &  H and K  \\

\hline        
   BD+00 2058A &  -0.55   & -0.57 &  0.18   & 0.29  \\
   BD+02 2541  &   ---    & -0.22 &  0.22   & -0.22  \\
   BD+06 2932  &  -0.58   & -0.63 &  0.28   & -0.15  \\
   BD+07 4841  &  -0.66   & -0.65 &  0.16   & 0.24  \\
   BD+11 2369  &  -0.15   & -0.19 &  0.14   & -0.12  \\
   BD+11 4725  &  -0.19   & -0.22 &  0.14   & -0.12  \\
   BD+18 3423  &  -0.66   & -0.66 &  0.13   & 0.10  \\
   BD-01 306   &  -0.25   & -0.30 &  0.21   & 0.00  \\
   BD-03 56    &  -0.28   & -0.28 &  0.22   & 0.05  \\
   BD-06 855   &  -0.61   & -0.65 &  0.20   & -0.16  \\
   BD-08 4501  &  -0.72   & -0.72 &  0.16   & 0.29  \\
   CD-33-3337  &  -0.50   & -0.46 &  0.08   & 0.33  \\
   CD-43 6810  &  -0.28   & -0.28 &  0.16   & -0.10  \\
   CD-57 1633  &  -0.55   & -0.59 &  0.09   & 0.00  \\
   G 24-25     &   ---    & -0.60 &  0.15   & 0.00   \\
   G 63-26     &  -0.24   & -0.20 &  ---    &  0.36  \\
   G 75-31     &  -0.65   & -0.67 &  0.16   &  0.15  \\
   HD 103723   &  -0.54   & -0.61 &  0.18   &  0.07  \\
   HD 105004   &  -0.52   & -0.50 &  0.09   &  0.05  \\
   HD 106516   &  -0.30   & -0.26 &  0.25   &  0.07  \\
   HD 111980   &  -0.28   & -0.33 &  0.22   &  0.15  \\
   HD 113679   &  -0.21   & -0.24 &  0.24   &  0.00  \\
   HD 120559   &  -0.17   & -0.17 &  0.22   &  -0.11  \\
   HD 121004   &  -0.17   & -0.21 &  0.23   &  -0.02  \\
   HD 132475   &  -0.28   & -0.32 &  0.19   &  0.19  \\
   HD 148816   &  -0.33   & -0.33 &  0.21   &  -0.05  \\
   HD 159482   &  -0.29   & -0.29 &  0.35   &  0.12  \\
   HD 163810   &  -0.64   & -0.69 &  0.19   &  -0.08  \\
   HD 177095   &  -0.25   & -0.25 &  0.13   &  -0.19  \\
   HD 179626   &  -0.39   & -0.39 &  0.23   &  0.04  \\
   HD 189558   &  -0.39   & -0.39 &  0.28   &  0.12  \\
   HD 193901   &  -0.76   & -0.80 &  0.23   &  0.00  \\
   HD 194598   &  -0.60   & -0.61 &  0.09   &  0.10  \\
   HD 199289   &  -0.21   & -0.27 &  0.21   &  0.07  \\
   HD 205650   &  ---     & -0.26 &  0.21   &  0.05  \\
   HD 219617   &  -0.60   & -0.57 &  0.23   &  0.24  \\
   HD 22879    &  -0.32   & -0.36 &  0.28   &  0.00  \\
   HD 284248   &  -0.68   & -0.68 &  0.14   &  0.34  \\
   HD 298986   &  -0.65   & -0.65 &  0.09   &  0.26  \\
   HD 3567     &  -0.65   & -0.66 &  0.10   &  0.20  \\
   HD 51754    &  -0.18   & -0.23 &  0.23   & -0.09  \\
   HD 59392    &  -0.60   & -0.67 &  0.15   &  0.37  \\
   HD 76932    &  -0.22   & -0.30 &  0.27   &  0.09  \\
   Wolf 615    &  -0.62   & -0.67 &  0.14   &  0.07  \\
   CD-61 282   &  -0.60   & -0.62 &  0.17   &  0.10  \\
\hline
\multicolumn{5}{c}{Archival spectra} \\
\hline
CD-45 3283      &  -0.71 & -0.71  & ---    & -0.10  \\
G 46-31         &  ---   & ---    & ---    & 0.00  \\   
G 87-13         &  -0.63 & -0.63  & 0.07   & 0.16  \\
G 98-53         &  -0.68 & -0.66  & 0.26   & 0.04  \\
G 114-42        &  -0.52 & -0.54  & 0.14   & 0.04  \\
G 161-73        &  -0.62 & -0.61  & 0.14   & 0.11  \\  
G 53-41         &  -0.58 & -0.58  & ---    & 0.16  \\
G 119-64        &  -0.69 & -0.69  & 0.17   & 0.31  \\  
G 05-36         &  -0.45  & -0.43   & 0.33 & 0.33 \\  
G 188-22        &  -0.36  & -0.38   & 0.29 & 0.31 \\  
HD 25704       &  -0.30  & -0.32   & 0.32   & 0.03 \\  

   \hline
   \hline
    \end{tabular}}
    \label{tab:Abonds}
\end{table}

\section{Discussion}

In this section, we examine whether aluminium abundances serve as a reliable discriminator between in situ and accreted halo populations and evaluate the strengths and limitations of this approach. Our analysis shows that the division between high- and low-aluminium stars is robust across the full metallicity range and that NLTE corrections enhance the clarity of this separation. We next discuss the implications of this result for population classification, nucleosynthesis, kinematic selections, and its relation to the Nissen \& Schuster classification.

\subsection{High-aluminium and low-aluminium stars}

\begin{figure}
    \centering
    \includegraphics[width=0.9\linewidth]{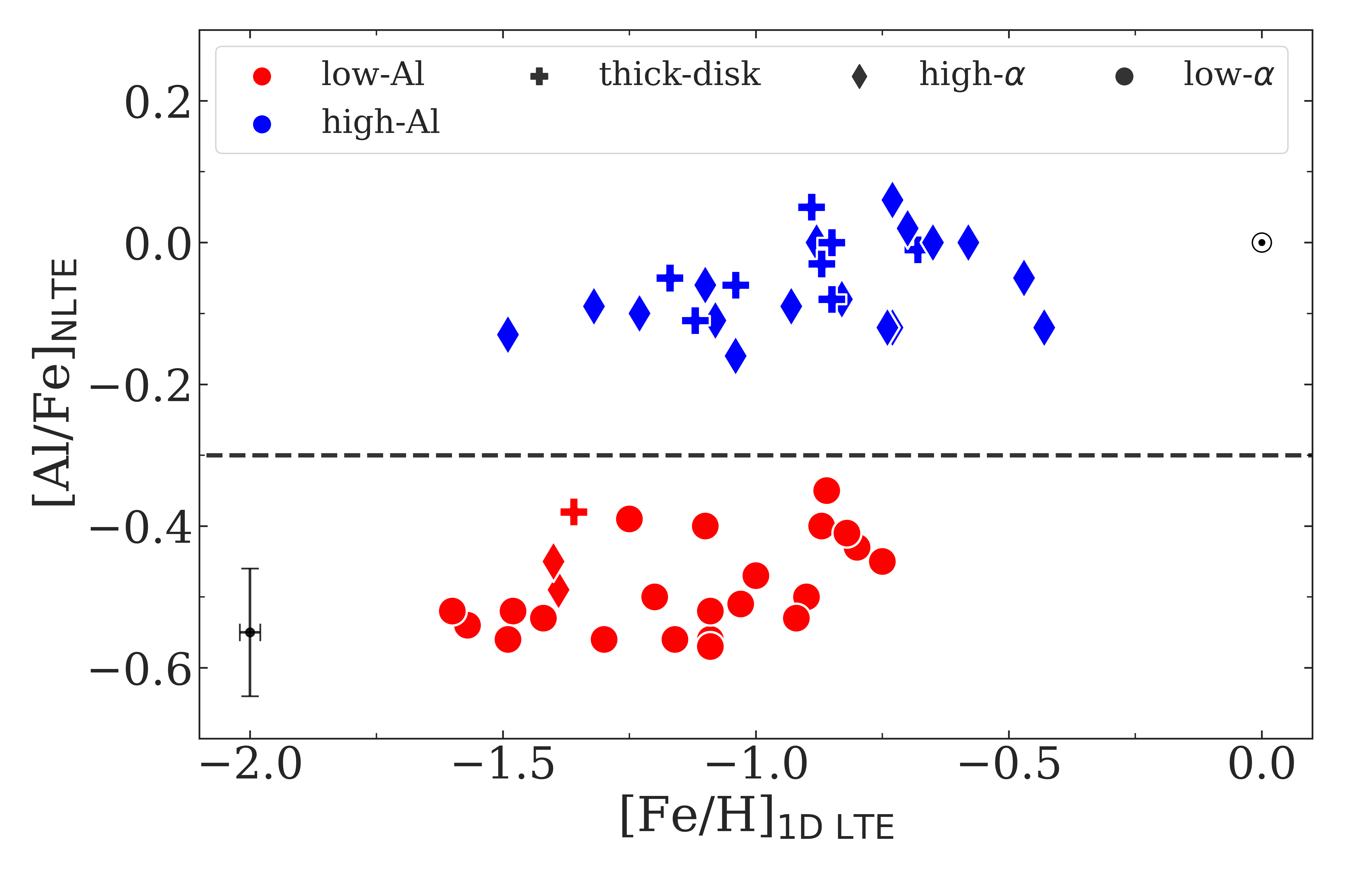}
    \includegraphics[width=0.9\linewidth]{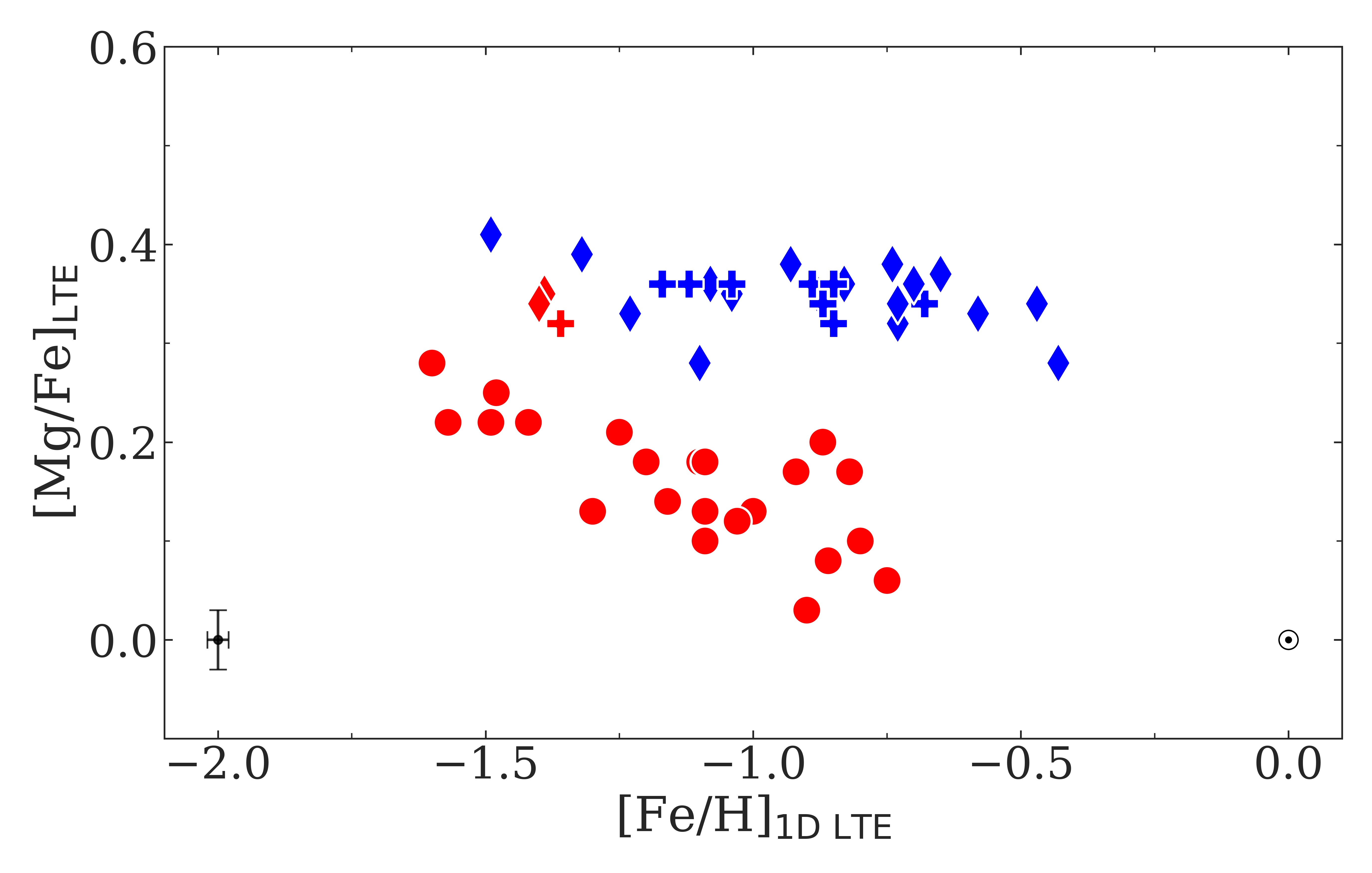}
    \caption{Top panel: Relation between [Al/Fe] and [Fe/H] for the stellar sample. The elemental abundances derived under NLTE assumptions for the Al I 3961 {\rm \AA} line. The data are colour-coded by high-Al and low-Al, with symbols following the classification of \citep{Nissen10}. Thick-disc stars are represented by crosses, high-$\alpha$ stars are shown as diamonds, and low-$\alpha$ stars as circles. Bottom panel: Distribution of the stellar sample in the [Mg/Fe]$_{\rm NLTE}$ versus [Fe/H]$_{\rm LTE}$, according to the low-Al and high-Al classification used in this paper, adopting the same symbols as the top panel.}
    \label{Al-classplot}
\end{figure}

In Fig.~\ref{Al-classplot} (top panel) we classify the sample into high- and low-Al stars using NLTE abundances from the Al\,I 3961\,\AA\ line.

A clear division emerges at [Al/Fe] = $-0.3$, which we set as the threshold separating high- and low-Al stars. This boundary distinguishes stars with systematically enhanced aluminium abundances, predominantly associated with the in situ halo and thick disc, from those with depleted values that trace the accreted population. The bottom panel situates this division in the [$\alpha$/Fe] context, showing [Mg/Fe] as a function of [Fe/H]. The high-Al stars overlap closely with the high-$\alpha$ sequence, while the low-Al stars coincide with the low-$\alpha$ sequence. In particular, at high $\rm[Fe/H]>-1.3$, the low-$\alpha$ and low-Al populations are identical. However, at low $\rm[Fe/H]<-1.3$, we identify two low-Al stars previously categorised as high-$\alpha$ (red diamonds) and one as a thick-disc star (red cross).
The separation of the two populations in [$\alpha$/Fe] diminishes at low [Fe/H], where both $\alpha$-elements and Fe mainly originate from core-collapse supernovae. In contrast, the separation in [Al/Fe] remains clear, underscoring the discriminating power of aluminium as a powerful tracer of accreted halo stars, especially at low [Fe/H].

This classification provides a clean division in chemical abundance space and offers an important advantage for disentangling halo populations whose kinematic signatures alone may be ambiguous, especially at lower energies. Aluminium can be a complementary tracer to $\alpha$-elements for chemical tagging of halo stars. However, as discussed in the following, the interpretation of aluminium abundances must be tempered by an understanding of their nucleosynthetic origin and limitations.

\subsection{Aluminium nucleosynthesis and limitations}

Aluminium is produced in massive stars through the burning of hydrostatic carbon and neon \citep{WW95}. It can act as a primary element, but also as a secondary element via $\alpha$-, n-, and p-captures on neutron-rich seeds such as $^{22}$Ne and $^{23}$Na \citep{Roederer21}. In the Milky Way, aluminium behaves mainly as a secondary element, with metallicity-dependent trends \citep{Kobayashi20}. Its yield also depends on both the stellar population and the star formation history across different environments.

\citet{Hasselquist24} investigated aluminium abundances in the Milky Way and its satellite galaxies using a two-process model, in which abundances are interpreted as the sum of a prompt component from core-collapse supernovae and a delayed contribution from Type~Ia supernovae. They report that [Al/Mg] residuals in the Large Magellanic Cloud and the Sagittarius dwarf spheroidal galaxy correlate more strongly with [(C+N)/H] than with [Mg/H], suggesting that C and N act as more direct tracers of the nucleosynthetic pathways responsible for aluminium.  

Their study shows that when the two-process model is calibrated on Milky Way stars, it systematically overpredicts aluminium abundances in satellite galaxies by 0.1–0.3\,dex, particularly at higher metallicities. This mismatch implies that aluminium enrichment in dwarf galaxies follows a different chemical evolution from that in the Milky Way. When the model is recalibrated using Sagittarius stars, the resulting dwarf-galaxy-trained model successfully reproduces aluminium abundances across a wide range of satellites. Although aluminium is a powerful tracer, its interpretation depends sensitively on the Galactic environment, and models must be tailored accordingly.  

Together, these considerations show both the power and the limitations of aluminium as a chemical tag. Although aluminium abundances provide a robust empirical discriminator between in situ and accreted populations, their nucleosynthetic interpretation is complex and requires careful consideration of stellar yields, NLTE effects, and environmental dependence.

\subsection{Kinematic selections}

\begin{figure*}
    \centering
    \includegraphics[width=0.9\linewidth]{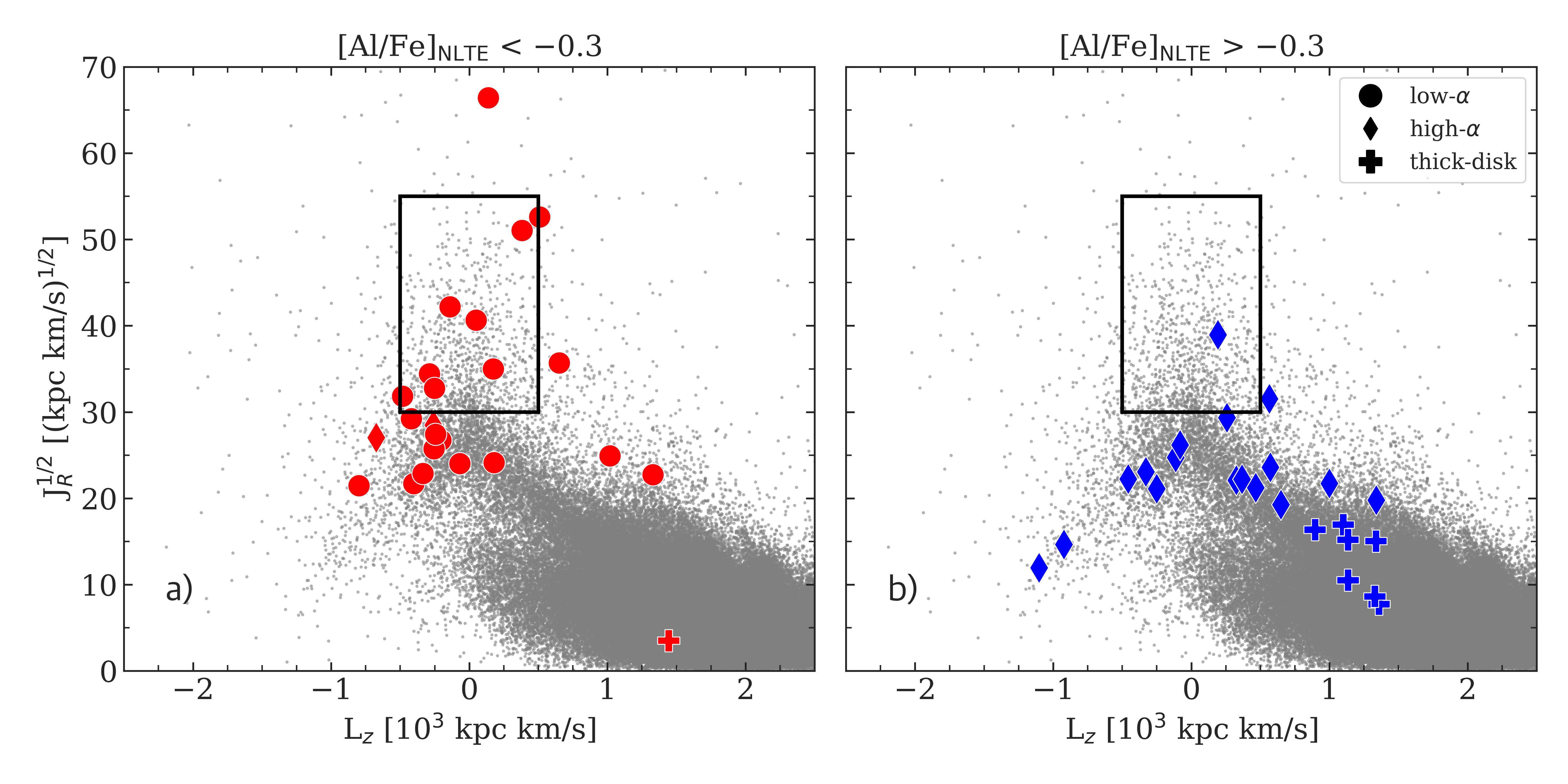}
    \caption{Relation between $\sqrt{J_R}$ and angular momentum, $L_z$, for the stellar sample. Panel (a): Low-Al stars. Panel (b): High-Al stars. The low-$\alpha$ stars are shown as filled circles, the high-$\alpha$ stars as diamonds, and the thick-disc stars are displayed as crosses in both panels. The black box marks the \cite{Feuillet22} selection scheme for the {\it Gaia}-Sausage-Enceladus. APOGEE DR17 data are plotted in the background.}
    \label{kin}
\end{figure*}

We calculated the energies, radial action angles, and angular momenta of the stars in our sample using the \texttt{galpy} package \citep{Bovy15}. Distances (distance\_gspphot) and proper motions were adopted from {\it Gaia} Data Release~3 \citep{Gaia16b,Gaia23j}, while radial velocities were taken from \citet{Nissen10,Nissen11}. We assumed the \texttt{MWPotential2014} Galactic potential model \citep{Bovy15} and applied the actionAngleStaeckel approximation \citep{Binney12,Bovy13} using a delta value of 0.4, which specifies the focal length of the assumed oblate potential. The typical uncertainties in the input parameters are 0.7\,pc in distance (i.e. a relative uncertainty of $\lesssim 1\%$ for our sample), 0.03\,mas\,yr$^{-1}$ in proper motion, and 0.3\,km\,s$^{-1}$ in radial velocity. Propagating these uncertainties through a Monte Carlo analysis, as described by \citet{Feuillet20}, yields typical uncertainties of $0.12 \times 10^3$\,kpc\,km\,s$^{-1}$ in angular momentum, $0.28 \times 10^4$\,km$^2$\,s$^{-2}$ in orbital energy, and 2 (kpc\,km\,s$^{-1}$)$^{1/2}$ in $\sqrt{J_R}$.  

The distribution of stars in the $\sqrt{J_R}$–$L_Z$ space is shown in Fig.~\ref{kin}. The selection proposed by \citet{Feuillet21},  described by \citet{Carrillo23} as one of the purest kinematic criteria to identify the {\it Gaia}–Sausage–Enceladus population, is confirmed to have a high degree of purity when applied to our low-$\alpha$ and low-aluminium stars. However, we also find that this selection misses a high fraction ($>$50\%) of low-Al stars at smaller $\sqrt{J_R}$ values, deeper in the Galactic potential, where the distinction between accreted and in situ stars becomes more ambiguous. 
This difficulty is exacerbated when adopting more realistic Galactic potentials that account for the influence of non-axisymmetric structures, such as the bar, which further complicates the kinematic separation of halo populations \citep{Dillamore25}.  

Our results suggest that a refined kinematic selection is possible when combined with aluminium abundances. Specifically, stars with [Al/Fe] $<-0.3$ are highly likely to belong to the {\it Gaia}–Sausage–Enceladus population, even when their kinematics fall outside the strict boundaries of proposed criteria. Within our dataset, low-aluminium stars are found in the region $-1000 < L_Z < 500$\,kpc\,km\,s$^{-1}$ and $20 < \sqrt{J_R} < 55$\,kpc\,km\,s$^{-1/2}$ when aluminium abundances are available. This suggests that extending existing kinematic selections to include stars with lower $J_R$ values can improve completeness without significantly sacrificing purity.  

We emphasise, however, that abundance-based selections should be calibrated on the data in the case of a survey, since the precise threshold in [Al/Fe] may vary between surveys depending on the adopted analysis and systematic differences in stellar parameters (see Sect. \ref{sect:MgMnAl}).

\subsection{[Mg/Mn] versus [Al/Fe] using the Al\,I resonance lines}
\label{sect:MgMnAl}

\begin{figure}
    \centering
    \includegraphics[width=0.9\linewidth]{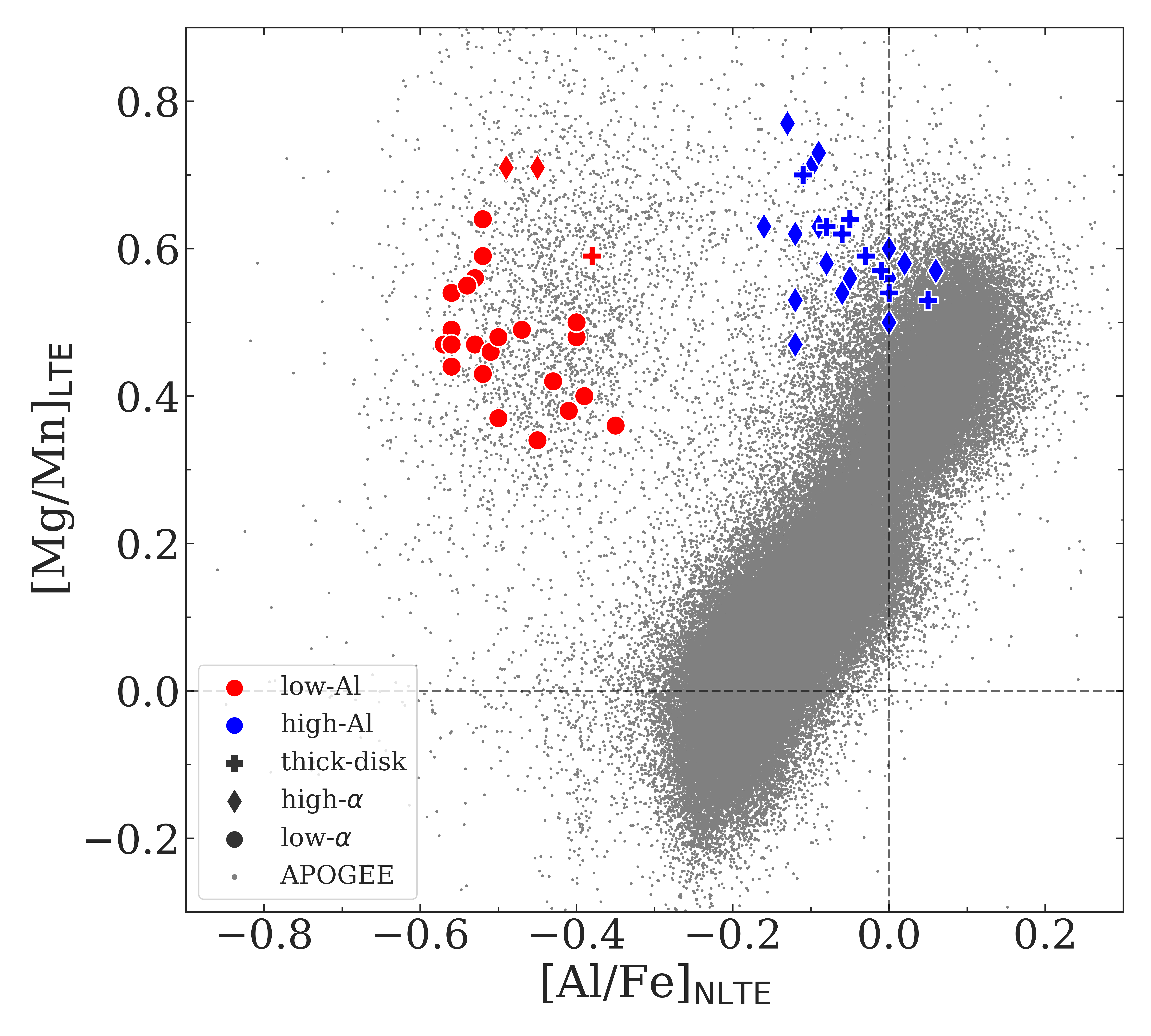}
    \caption{Distribution of stars in the [Mg/Mn] versus [Al/Fe] parameter space. The background grey points show APOGEE DR17 giant stars (log$g$<3.0) for reference. Our sample is overplotted with the same symbols as in Fig. \ref{Al-classplot}.}
    \label{MgMnAlFeplot}
\end{figure}

An important step is to place our [Al/Fe] measurements in context with other elemental abundances and independent datasets. In Fig.~\ref{MgMnAlFeplot},  we show our sample of the Nissen \& Schuster stars in the [Mg/Mn] versus [Al/Fe] plane. This diagnostic has recently been shown to be a powerful tool for disentangling accreted populations in situ, since magnesium and manganese probe different nucleosynthetic channels, while aluminium provides an additional constraint \citet{Hawkins15}. Our sample consistently falls into the expected loci of high and low$\alpha$ populations, supporting the robustness of the Al-based classification.  

When comparing our aluminium abundances with those derived from APOGEE \citep{Majewski17APOGEE}, based on the H-band Al\,I lines near 1.5\,$\mu$m, we find a systematic offset of $\sim+0.25$\,dex. Most of this offset can be explained by the NLTE correction of the Al H-band lines. As discussed in Section. 4.2.1 of \citet{Feltzing23}, the NLTE corrections for the Al lines in the H-band are expected to be about -0.2 dex in [Al/H] for a giant star such as Arcturus, based on the calculations by \citet{NordlanderLind17}. Since our APOGEE subsample consists exclusively of giants, we adopted this value as a representative correction and applied it to the APOGEE [Al/Fe] abundances. We corrected the [Al/Fe] values by -0.2 dex, as \citet{Feltzing23} argue that NLTE effects on [Fe/H] should be negligible for these stars. 

For the magnesium and manganese corrections, \citet{Feltzing23} discuss that for giant stars the NLTE correction for the APOGEE lines corresponds to about $-0.35$ dex in [Mg/Mn]. We verified this value using the MPIA web interface\footnote{https://nlte.mpia.de/gui-siuAC\_secE.php}, which provides consistent results for the H-band lines. For dwarf stars, we considered the corrections derived from the optical lines for Mg and Mn, as adopted in \citet{Nissen10}. In this case, the NLTE corrections are negligible ($\sim0.0$~dex) for [Mg/Fe] and about $+0.10$ dex for [Mn/Fe]. Taken together, this implies an NLTE correction of roughly $-0.10$ dex in [Mg/Mn] for dwarf stars, $\sim0.25$~dex smaller than the value found for giants. Since these are very rough estimates, and we are not focusing on the Mg and Mn lines, we chose not to apply these corrections to the $y$-axis of Fig.~\ref{MgMnAlFeplot}, and instead present only [Mg/Mn]$_{\rm LTE}$. A more complete treatment of NLTE effects, including $\langle 3D \rangle$ NLTE corrections for Mg, Mn, Al, and Fe across a theoretical grid, is underway and will be presented by Ernandes et al. (in preparation).

\subsection{Comments on individual stars}

In the new classification of low-Al and high-Al stars, some objects show discrepancies relative to the Nissen \& Schuster classification of low- and high-$\alpha$~stars.
The stars BD+07 4841, CD-33 3337, and G 24-25 exhibit unexpected aluminium abundances considering their previous classification, as shown in Fig.~\ref{Al-classplot}.

\begin{figure}
    \centering
    \includegraphics[width=0.9\linewidth]{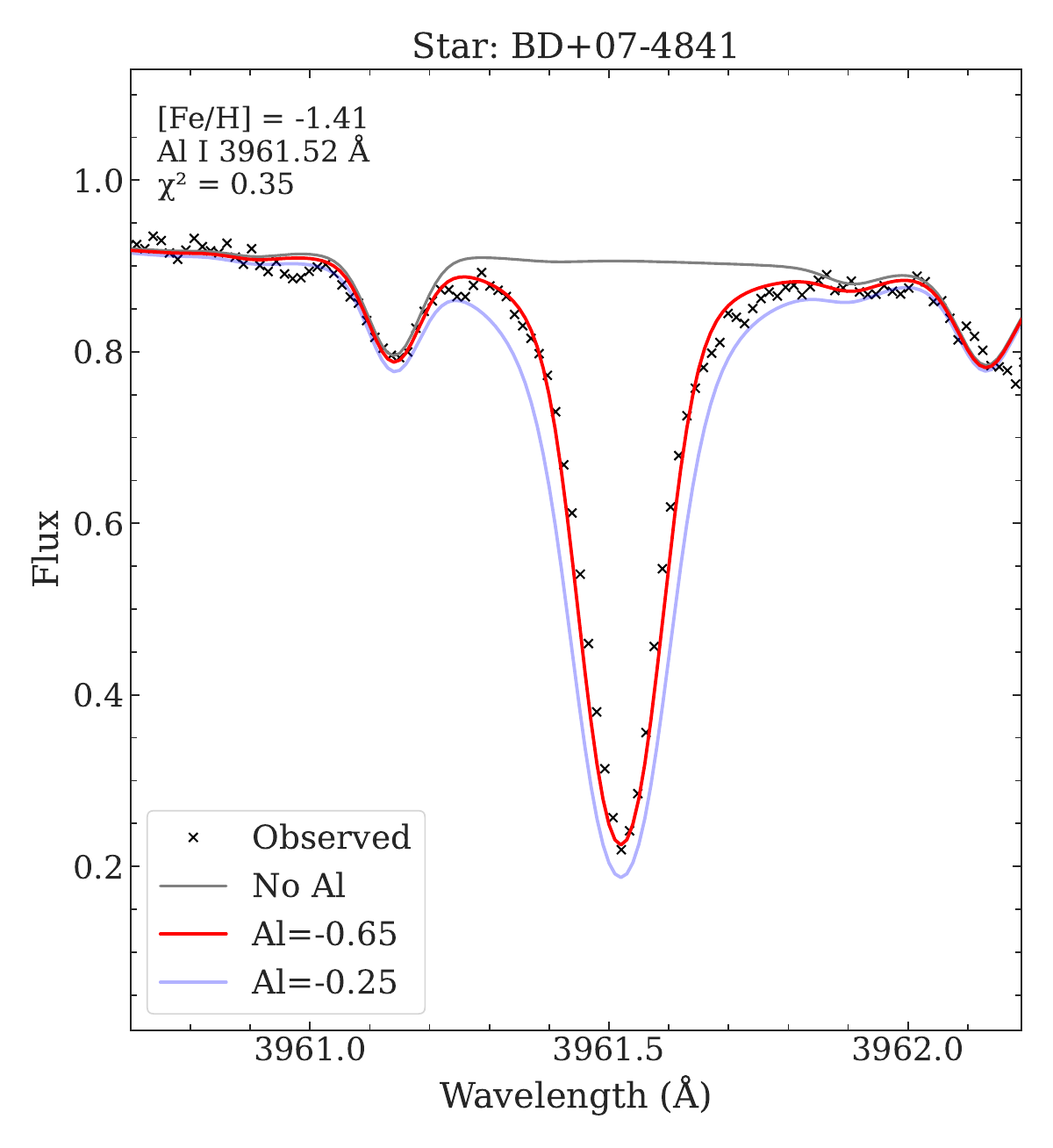}
    \includegraphics[width=0.9\linewidth]{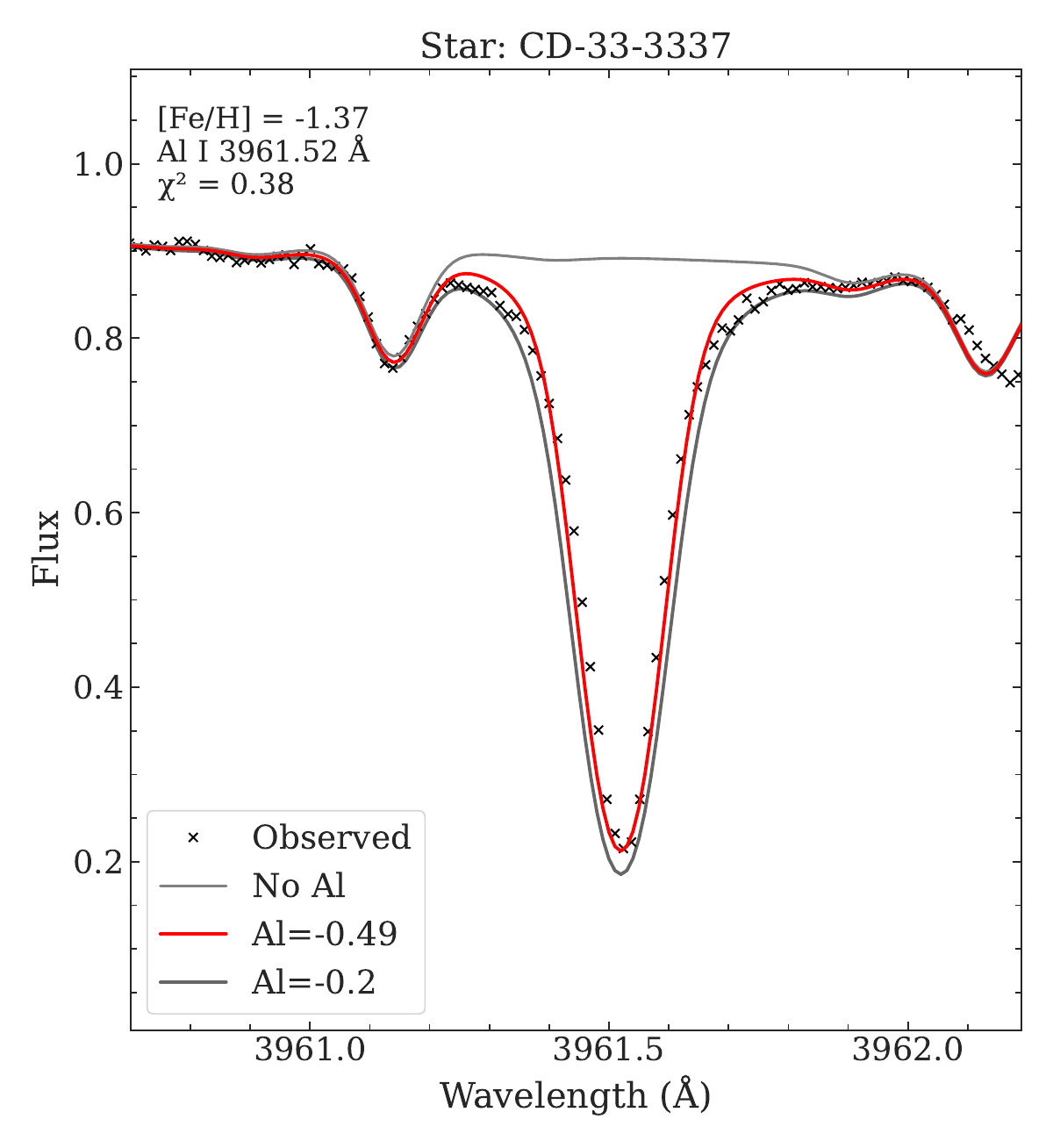}    
    \caption{Line profile fits for the Al I LTE at the 3961{\rm \AA} line in two stars that show atypical abundance patterns for their assigned class in \cite{Nissen11}. Panel (a): BD+07 4841, a high$-\alpha$ star with low [Al/Fe]; and panel (b): CD-33-3337, a thick-disc (TD) star with unusually low [Al/Fe] compared to typical TD trends. Observed spectra are shown in black, and the synthetic fits in colour for different [Al/Fe] values.}
    
    \label{fig:synHA}
\end{figure}

\begin{itemize}
    
    \item BD+07 4841: This star was classified as a high-$\alpha$ star by Nissen \& Schuster (2010). However, its aluminium abundance does not follow the typical pattern of high-$\alpha$ stars, as is also evident from the line fitting shown in Fig.~\ref{fig:synHA}.
    
    \item CD-33 3337: This star was classified by \cite{Nissen10} as a thick-disc star based on its kinematics. However, its aluminium abundance differs significantly from that expected for thick-disc stars, as shown in the [Al/Fe] versus [Fe/H] plot in Fig.~\ref{Al-classplot}. Additionally, its [Mg/Mn] and [Al/Fe] ratios correspond to a low-aluminium disc population \citep{Feltzing23}, as shown in Fig.~\ref{MgMnAlFeplot}. The derived abundances are confirmed by the spectral line fitting in Fig.~\ref{fig:synHA}. In the $En$–$L_z$ diagram, the star remains within the disc region.
    
    \item G 24-25: This star is classified as a high-$\alpha$ star by \cite{Nissen10}, and in SIMBAD, it is identified as both a spectroscopic binary (SB1) and a CEMP-s, features that are evident in its spectrum shown in Fig.~\ref{spectra}.
\end{itemize}

Because these stars have [Fe/H] $\sim -$1.5, a region where the two populations are difficult to distinguish in the [$\alpha$/Fe] plot, this may explain the classification difference for BD+07 4841. This may also apply to G 24-25, or it may simply reflect the fact that CEMP-s stars display unusual abundance patterns, which could lead to its misclassification.
As the classifications based on kinematics and [Al/Fe] abundance are inconsistent for CD-33 3337, further investigation is required to determine its origin. 

\section{Conclusions}

We have derived aluminium abundances for a carefully selected subsample of halo stars from the Nissen \& Schuster dataset, combining new VLT/UVES observations with archival spectra. Elemental abundances were derived from the Al\,I 3944 and 3961\,{\rm \AA} resonance lines using line-profile fitting under both LTE and NLTE assumptions. Our main conclusions are as follows.  

First, the separation between accreted and in situ halo populations is clear in aluminium abundances, even at low $\rm[Fe/H]\approx-1.5$. This distinction is already visible in LTE but becomes significantly enhanced when NLTE corrections are applied. We propose a classification threshold at [Al/Fe]\,=\,$-0.3$ to distinguish between high- and low-aluminium stars, which correlates to in situ and accreted stars, respectively.

Second, aluminium abundances provide an effective chemical diagnostic that complements $\alpha$-element abundances. In particular, aluminium offers a cleaner separation than [$\alpha$/Fe] at low metallicities, thereby improving the classical Nissen \& Schuster classification, especially when NLTE corrections are applied.  

Third, while aluminium is a powerful empirical discriminator, its nucleosynthetic interpretation is complex. Aluminium can behave as either a primary or secondary element, depending on the underlying stellar populations and the chemical enrichment history \citep{WW95,Kobayashi20}. Comparisons with dwarf galaxies confirm that chemical evolution models calibrated on the Milky Way systematically overpredict aluminium in external systems, underlining the need for environment-specific yields and models \citep{Hasselquist24}.  

Finally, we emphasise the importance of combining elemental abundances with kinematic information. Although some kinematic selection can provide high-purity samples of accreted stars, they can miss parts of the population, particularly at low energies, which can introduce biases. Aluminium abundances provide an independent and complementary way to probe, providing a robust means to distinguish between accreted and in situ halo populations in the metal-poor regime. Aluminium has also been used to investigate internal variations in star formation history within the {\it Gaia}-Sausage-Enceladus progenitor galaxy, as shown by \cite{Asa25}.

Overall, our results confirm that aluminium abundances, when measured with high-quality spectra and analysed with NLTE corrections, constitute a powerful tool to trace the origin of halo stars. They provide a reliable chemical tag for accreted populations, opening new possibilities for disentangling the complex assembly history of the Milky Way.

%
%

\section*{Acknowledgements}
H.E., S.F., and D.F. were supported by a project grant from the Knut and
Alice Wallenberg Foundation (KAW 2020.0061 Galactic Time
Machine, PI Feltzing). H.E. is supported by foundations managed by The Royal Swedish Academy of Sciences. and supported by The Royal Physiographic Society in Lund (reference 162011). This project was supported by funds from the Crafoord Foundation (reference 20230890).
D.F.~acknowledges funding from the Swedish Research Council grant 2022-03274.
\'{A}.S.~acknowledges funding from the European Research Council (ERC) under the European Union’s Horizon 2020 research and innovation programme (grant agreement No. 101117455).

%
%

\bibliographystyle{aa}
\bibliography{stars}

\begin{appendix} 

\section{Radial velocity and stellar parameters}
We present the radial velocities derived following the methodology in Section 2. We note that HD 111980 is a known spectroscopic binary, SB1, as listed in Simbad.

\begin{table}[]
    \centering
\caption{Measured radial velocities for the stellar sample, corrected to the heliocentric rest frame. }
    \begin{tabular}{lll}
\hline
\hline
 Star &  OBs &  Radial velocity  \\
 & & [kms$^{-1}$] \\
\hline
   BD+00 2058A  &  1  &    +83.28  \\
   BD+02 2541   &  2  & $-$153.80  \\
   BD+06 2932   &  1  &   +144.31  \\
   BD+07 4841   &  1  &   +234.19  \\
   BD+11 2369   &  1  &  $-$28.37  \\
   BD+11 4725   &  1  &   +197.03  \\
   BD+18 3423   &  1  &   +240.81  \\
   BD-01 306    &  1  &  $-$28.91  \\
   BD-03 56     &  1  &    +28.19  \\
   BD-06 855    &  1  & $-$295.46  \\
   BD-08 4501   &  1  &  $-$84.58  \\
   CD-33 3337   &  1  &  $-$74.30  \\
   CD-43 6810   &  1  & $-$166.23  \\
   CD-57 1633   &  4  & $-$261.02  \\
   CD-61 282    &  1  & $-$220.80  \\
   G 24-25      &  1  &   +303.93  \\
   G 63-26      &  4  &  $-$58.04  \\
   G 75-31      &  1  &  $-$58.00  \\
   HD 103723    &  1  & $-$168.80  \\
   HD 105004    &  1  & $-$122.12  \\
   HD 106516    &  2  &  $-$12.13  \\
   HD 111980*   &  3  & $-$150.02  \\
   HD 113679    &  1  & $-$157.84  \\
   HD 120559    &  1  &  $-$17.70  \\
   HD 121004    &  1  & $-$245.07  \\
   HD 132475    &  4  & $-$176.66  \\
   HD 148816    &  1  &    +47.73  \\
   HD 159482    &  2  &   +135.28  \\
   HD 163810    &  2  & $-$185.17  \\
   HD 177095    &  1  &  $-$90.72  \\
   HD 179626    &  1  &    +64.74  \\
   HD 189558    &  1  &    +12.34  \\
   HD 193901    &  2  &   +171.19  \\
   HD 194598    &  3  &   +247.02  \\
   HD 199289    &  1  &     +5.94  \\
   HD 205650    &  1  &   +102.38  \\
   HD 219617    &  2  &  $-$13.91  \\
   HD 22879     &  1  & $-$120.49  \\
   HD 284248    &  2  & $-$339.08  \\
   HD 298986    &  1  & $-$198.63  \\
   HD 3567      &  2  &    +47.43  \\
   HD 51754     &  1  &    +94.05  \\
   HD 59392     &  2  & $-$268.51  \\
   HD 76932     &  1  & $-$120.18  \\
   Wolf 615     &  2  & $-$170.67  \\
\hline
\multicolumn{3}{c}{Archival spectra} \\
\hline
  CD-45-3283     &  1  & $-$309.36  \\
  G 46-31        &  3  & $-$227.19  \\   
  G 87-13        &  1  & $-$206.46  \\
  G 98-53        &  1  & $-$144.08  \\
  G 114-42       &  2  & +86.300  \\
  G 161-73       &  1  & $-$121.04   \\  
  G 53-41        &  3  & $-$88.30  \\
  G 119-64       &  1  & +196.10  \\  
G 05-36          & 1  &  +9.88  \\     
G 188-22         & 1  &  +94.77  \\     
HD 25704        & 9  &  $-$56.130 \\     
\hline
\hline
    \end{tabular}
    \tablefoot{$*$ Known spectroscopic binaries.}
    \label{tab:rv}
\end{table}

Table \ref{tab:pars} displays Effective temperature ($T_{\rm eff}$), surface gravity (log$g$), microturbulence (v$_{\rm mic}$), and [Fe/H], 1D and 3D for each star, along with their original classification (high$-\alpha$, low$-\alpha$, or thick-disc) and the source of the stellar parameters, as “NSflag” with “NS24” for \cite{Nissen24} and “NS11” for \cite{Nissen11}. In the last column, the new classification based on the Aluminium elemental abundance in the current paper is presented.

\begin{table*}[]
    \centering
\caption{Stellar parameters for the stars analysed in this work.}
\resizebox{0.93\textwidth}{!}{
\begin{tabular}{lllccccccccc}
\hline
\hline
 Star-ID &  \it{Gaia} DR3 ID & N24-ID & T$_{eff}$ & log$g$ & v$_{mic}$ & \multicolumn{2}{c}{[Fe/H]} & NSflag &NS Class & Al Class \\
  &   &   &   & &  &  1D  & 3D  &  &  & & \\
 &   &   & [K] & &[km s$^{-1}$] &   & &  &  & & \\
\hline
BD+00 2058A  & 3085891537839267328 & G 112-43   & 6209 & 4.02 & 1.17 & -1.27 & -1.28 & NS24 & low-$\alpha$  & low-Al \\
BD+02 2541   & 3700341138433848832 & G 13-38    & 5263 & 4.54 & 0.90 & -0.88 &  ---  & NS11 & high-$\alpha$ & high-Al  \\
BD+06 2932   & 1159108770069883136 & G 66-22    & 5297 & 4.46 & 0.78 & -0.88 & -0.90 & NS24 & low-$\alpha$	& low-Al  \\
BD+07 4841   & 2722849325377392384 & G 18-39    & 6112 & 4.23 & 1.44 & -1.41 & -1.39 & NS24 & high-$\alpha$ & low-Al  \\
BD+11 2369   & 3916257626962267136 & G 57-07    & 5755 & 4.33 & 0.99 & -0.48 & -0.51 & NS24 & high-$\alpha$ & high-Al  \\
BD+11 4725   & 2728314787225296000 & G 18-28    & 5443 & 4.49 & 0.88 & -0.85 & -0.86 & NS24 & high-$\alpha$ & high-Al  \\
BD+18 3423   & 4550762289589343488 & G 170-56   & 6112 & 4.11 & 1.39 & -0.94 & -0.93 & NS24 & low-$\alpha$  & low-Al  \\
BD-01 306    & 2494759795723713408 & G 159-50   & 5713 & 4.44 & 1.03 & -0.94 & -0.96 & NS24 & high-$\alpha$ & high-Al  \\
BD-03 56     & 2540830947835865728 & G 31-55    & 5731 & 4.35 & 1.26 & -1.12 & -1.12 & NS24 & high-$\alpha$ & high-Al  \\
BD-06 855    & 3202470247468181632 & G 82-05    & 5338 & 4.51 & 0.80 & -0.78 & -0.81 & NS24 & low-$\alpha$ & low-Al  \\
BD-08 4501   & 4165370682239910144 & G 20-15    & 6162 & 4.32 & 1.50 & -1.5  & -1.49 & NS24 & low-$\alpha$ & low-Al  \\
CD-33 3337   & 5579891291853909504 & CD-33 3337 & 6112 & 3.86 & 1.56 & -1.37 & -1.37 & NS24 & TD & low-Al   \\
CD-43 6810   & 5388634989413677440 & CD-43 6810 & 6059 & 4.32 & 1.24 & -0.44 & -0.44 & NS24 & high-$\alpha$ & high-Al   \\
CD-57 1633   & 5486881507314450816 & CD-57 1633 & 5981 & 4.29 & 1.08 & -0.91 & -0.93 & NS24 & low-$\alpha$ & low-Al   \\
G 24-25      & 4228176122142169600 & G 24-25    & 5828 & 3.86 & 1.20 & -1.40 & ---   & NS11 & high-$\alpha$ & low-Al   \\
G 63-26      & 3939515154043771392 & G 63-26    & 6175 & 4.17 & 1.65 & -1.58 & -1.58 & NS24 & high-$\alpha$ & ---   \\
G 75-31      & 2502689198705422848 & G 75-31    & 6135 & 4.02 & 1.28 & -1.04 & -1.05 & NS24 & low-$\alpha$	& low-Al   \\
HD 103723    & 3494135429225414016 & HD 103723  & 6050 & 4.20 & 1.11 & -0.81 & -0.82 & NS24 & low-$\alpha$	& low-Al   \\
HD 105004    & 3486853672952159872 & HD 105004  & 5852 & 4.35 & 1.09 & -0.83 & -0.84 & NS24 & low-$\alpha$	& low-Al  \\
HD 106516    & 3581001280225616128 & HD 106516  & 6327 & 4.43 & 1.18 & -0.69 & -0.71 & NS24 & TD &	high-Al   \\
HD 111980    & 3510294882898890880 & HD 111980  & 5878 & 3.98 & 1.39 & -1.09 & -1.08 & NS24 & high-$\alpha$ & high-Al   \\
HD 113679    & 6141991322784963072 & HD 113679  & 5761 & 4.05 & 1.37 & -0.66 & -0.65 & NS24 & high-$\alpha$ & high-Al   \\
HD 120559    & 5871764831114179968 & HD 120559  & 5486 & 4.58 & 1.05 & -0.91 & -0.90 & NS24 & TD & high-Al   \\
HD 121004    & 6095425184286680064 & HD 121004  & 5755 & 4.43 & 1.16 & -0.71 & -0.70 & NS24 & high-$\alpha$ & high-Al   \\
HD 132475    & 6232043867720079616 & HD 132475  & 5750 & 3.77 & 1.37 & -1.51 & -1.51 & NS24 & high-$\alpha$ & high-Al   \\
HD 148816    & 4433711099194059904 & HD 148816  & 5923 & 4.17 & 1.33 & -0.74 & -0.74 & NS24 & high-$\alpha$ & high-Al   \\
HD 159482    & 4485972467412021376 & HD 159482  & 5829 & 4.37 & 1.21 & -0.74 & -0.73 & NS24 & high-$\alpha$ & high-Al   \\
HD 163810    & 4150413750729448320 & HD 163810  & 5592 & 4.61 & 1.17 & -1.22 & -1.19 & NS24 & low-$\alpha$ & low-Al   \\
HD 177095    & 4082356695359119616 & HD 177095  & 5349 & 4.39 & 0.90 & -0.74 & ---   & NS11 & high-$\alpha$ & high-Al   \\
HD 179626    & 4263841431819084032 & HD 179626  & 5925 & 4.14 & 1.49 & -1.06 & -1.04 & NS24 & high-$\alpha$ & high-Al   \\
HD 189558    & 4189078626829509248 & HD 189558  & 5707 & 3.83 & 1.29 & -1.14 & -1.13 & NS24 & TD &	high-Al   \\
HD 193901    & 6859076107589173120 & HD 193901  & 5729 & 4.43 & 1.31 & -1.11 & -1.08 & NS24 & low-$\alpha$ & low-Al   \\
HD 194598    & 1752459532807704704 & HD 194598  & 6018 & 4.34 & 1.40 & -1.11 & -1.08 & NS24 & low-$\alpha$ & low-Al   \\
HD 199289    & 6481426917515214464 & HD 199289  & 5915 & 4.30  & 1.21 & -1.05 & -1.04 & NS24 & TD & high-Al   \\
HD 205650    & 6810862213470371072 & HD 205650  & 5793 & 4.35 & 1.17 & -1.19 & -1.17 & NS24 & TD & high-Al   \\
HD 219617    & 4263841431819084032 & HD 219617  & 5983 & 4.28 & 1.42 & -1.46 & -1.44 & NS24 & high-$\alpha$ & high-Al   \\
HD 22879     & 3250489115709122560 & HD 22879   & 5859 & 4.29 & 1.20 & -0.86 & -0.85 & NS24 & TD & high-Al   \\
HD 284248    & 52624073910834176   & HD 284248  & 6271 & 4.21 & 1.51 & -1.59 & -1.59 & NS24 & low-$\alpha$ & low-Al   \\
HD 298986    & 5358179685572496384 & CD-51 4628 & 6296 & 4.29 & 1.31 & -1.32 & -1.34 & NS24 & low-$\alpha$ & low-Al   \\
HD 3567      & 2427069874188580480 & HD 3567    & 6180 & 4.01 & 1.40 & -1.17 & -1.18 & NS24 & low-$\alpha$ & low-Al   \\
HD 51754     & 3112836479029139200 & HD 51754   & 5857 & 4.35 & 1.30 & -0.58 & -0.57 & NS24 & high-$\alpha$ & high-Al   \\
HD 59392     & 5586241315104190848 & HD 59392   & 6137 & 3.88 & 1.73 & -1.62 & -1.62 & NS24 & low-$\alpha$ & low-Al   \\
HD 76932     & 5730371484020807808 & HD 76932   & 5977 & 4.17 & 1.30 & -0.87 & -0.87 & NS24 & TD & high-Al   \\
Wolf 615     & 4412702764879270656 & G 16-20    & 5625 & 3.64 & 1.50 & -1.42 & ---   & NS11 & low-$\alpha$ & low-Al   \\
CD-61 282    & 4715919175280799616 & CD-61 282 & 5869 & 4.34 & 1.19 & -1.25 & -1.23 & NS24 & --- & low-Al   \\
\hline
\multicolumn{11}{c}{Archival spectra} \\ 
\hline
CD-45 3283    & 5510893810476230144 & CD-45 3283  & 5685 & 4.61 & 0.95 & -0.93 & -0.93 & NS24 & low-$\alpha$ &	---  \\
G 46-31       & 3845429561802456704 & G 46-31     & 6017 & 4.29 & 1.30 & -0.83 & -0.82 & NS24 & low-$\alpha$ &	--- \\   
G 87-13       & 940306024863622400  & G 87-13     & 6217 & 4.11 & 1.42 & -1.11 & -1.09 & NS24 & low-$\alpha$ &	low-Al \\
G 98-53       & 3440045641893430912 & G 98-53     & 5954 & 4.26 & 1.20 & -0.89 & -0.87 & NS24 & low-$\alpha$ &	low-Al \\
G 114-42      & 5759671677898613632 & G 114-42    & 5721 & 4.40 & 1.19 & -1.12 & -1.10 & NS24 & low-$\alpha$ &	low-Al \\
G 161-73      & 3822808140853060224 & G 161-73    & 6108 & 3.99 & 1.26 & -1.01 & -1.04 & NS24 & low-$\alpha$ &	low-Al \\  
G 53-41       & 3855475490307708032 & G 53-41     & 5956 & 4.34 & 1.35 & -1.21 & -1.16 & NS24 & low-$\alpha$ &	--- \\
G 119-64      & 761871677268717952  & G 119-64    & 6333 & 4.14 & 1.40 & -1.50 & -1.52 & NS24 & low-$\alpha$ &	low-Al \\  
G 05-36   &  68752844338431488   & G 05-36   & 6139   & 4.22   &  1.29 & -1.25 & -1.24  & NS24 & high-$\alpha$ & high-Al \\     
G 188-22  &  1800518563283501440 & G 188-22  & 6116   & 4.20   &  1.42 & -1.33 & -1.32  & NS24 & high-$\alpha$ & high-Al \\     
HD 25704       &  4682867168555426944 & HD 25704  & 5974   & 4.30  & 1.30 & -0.86 & -0.85 & NS24  & TD	& high-Al  \\   

\hline
\hline
    \end{tabular}
    }
    \tablefoot{“NS24” for \cite{Nissen24} and “NS11” for \cite{Nissen11}.}
    \label{tab:pars}
\end{table*}

\section{NLTE corrections}

\begin{figure}
    \centering
    \includegraphics[width=0.85\linewidth]{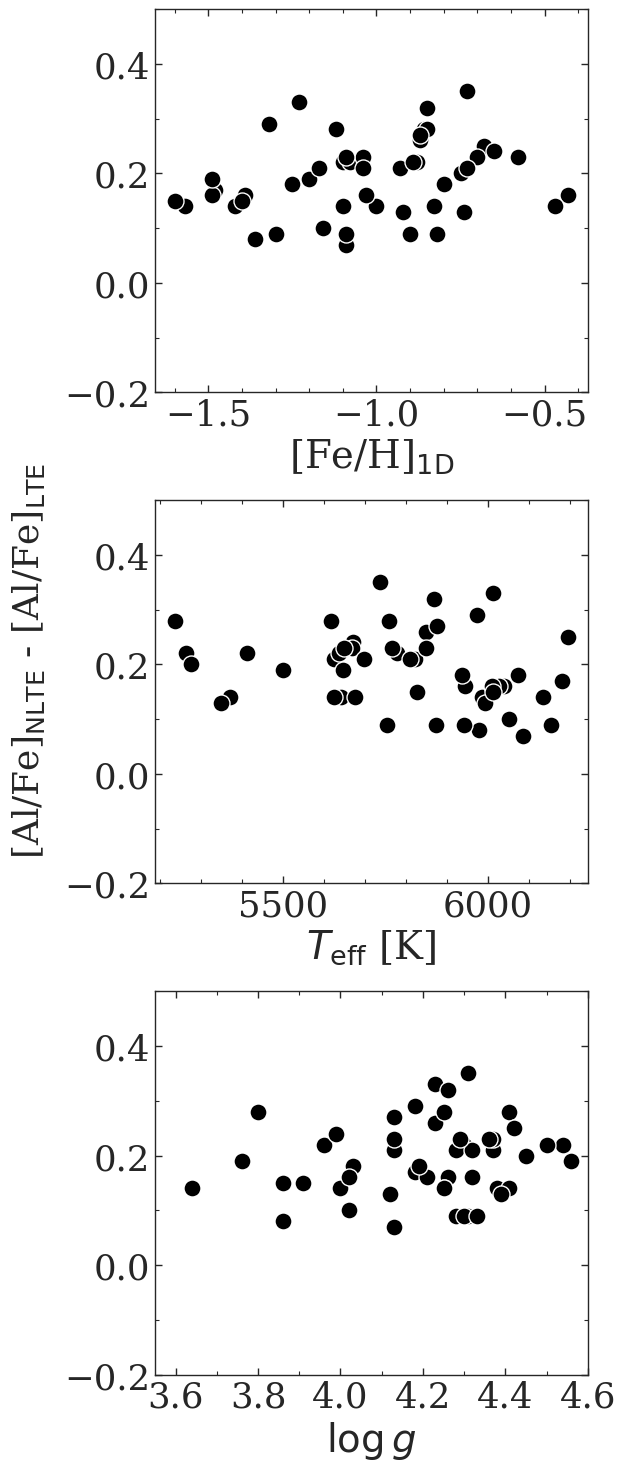}
    \caption{NLTE corrections for the Al I 3961 {\rm \AA} line as a function of stellar parameters. The three panels show the NLTE correction as a function of [Fe/H], $T_{\rm eff}$ and log $g$, respectively.}
    \label{stellar-plotnlte}
\end{figure}

We present the NLTE corrections applied to the Al I 3961 {\rm \AA} line for this sample as a function of [Fe/H]$_{\rm LTE}$, $T_{\rm eff}$,  and log $g$. As seen in Fig. \ref{stellar-plotnlte}, the corrections are $\sim$+0.2 dex in [Al/Fe] for all stars in our sample. We do not observe any significant dependency with stellar parameters.

\end{appendix}

\end{document}